\documentclass{aa}
\usepackage{psfig}
\usepackage{times}

\begin{document}

\addtolength{\topmargin}{20mm}

\thesaurus{01(11.01.2; 11.02.2 Mrk 501; 13.25.2)}

\def\nh{$N_{\rm H}$\ } 
\def\arx{$\alpha_{\rm {rx}}\,$}
\def\aro{$\alpha_{\rm {ro}}\,$}
\def\aox{$\alpha_{\rm {ox}}\,$}
\def\ax{$\alpha_{\rm {x}}\,$}
\def\am{$\langle\alpha\rangle\,$} 
\def\del{$\delta_{\rm IC}\,$}
\def\ch{$C_{\rm{H}}\,$}
\def\cs{$C_{\rm{S}}\,$}
\def\Ao{A_{\rm opt}}
\def\Az{A_{\rm z}}
\def\Ax{A_{\rm x}}
\def\P1{Paper I}

\title{Markarian 501 in X-ray bright state -- RXTE observations}
\author{ G. Lamer\inst{1} 
 \and S.J. Wagner\inst{2,3} }

\offprints{G. Lamer, Department of Physics and Astronomy, 
           University of Southampton,
           Highfield, Southampton, SO17 1BJ, England }

\institute{Institut f\"ur Astronomie und Astrophysik, Abt. Astronomie,
           Universit\"at T\"ubingen, 
           Waldh\"auserstr. 64, 
           D-72076 T\"ubingen, Germany\\
           e-mail: lamer@astro.uni-tuebingen.de
\and       Landessternwarte Heidelberg-K\"onigstuhl,
           D-69117 Heidelberg, Germany
\and       Mount Stromlo \& Siding Spring Observatories,
           Private Bag, Weston Creek PO,               
           ACT 2611, Australia }

\date{Received  ; accepted }

\maketitle

\begin{abstract}

Mrk 501 has been in a state of very high flux in  X-rays and VHE $\gamma$-rays
during 1997. In July 1997 near its hitherto maximum X-ray brightness 
intense multifrequency observations of Mrk 501
have been performed at radio, near infrared, optical, X-ray, 
and VHE $\gamma$-ray frequencies.

Here we report on {\em Rossi X-ray Timing Explorer} (RXTE) observations 
carried out in 1997 between July 11 and July 16. 
The X-ray spectrum has been well detected up to 100 keV and is best
described by a broken power law with 
spectral indices  $\alpha_1=0.70$ and $\alpha_2=0.94$ below
and above the break energy of $E_{\rm break}=5.8\; {\rm keV}$. 

The X-ray flux from Mrk 501 declined and flared by  $\sim 30\%$ 
within about 3 days each, showing an unusual 
anti-correlation between flux and spectral hardening of
both power law components. The break energy remained constant.

The observed broad band and X-ray spectra confirm recent observations, that
the synchrotron component in Mrk 501 can extend up to  energies of
100 keV with the maximum power being emitted at hard X-rays.
On longer time-scales  the cut-off frequency of the synchrotron spectrum
changes by more than two orders of magnitude.     

\keywords{ Galaxies: active -- BL Lacertae objects: individual: Mrk 501
-- X-rays: galaxies}
 
\end{abstract}

\section{Introduction}
\label{intro}

\begin{figure}[htb]
\par{\centerline{\psfig{figure=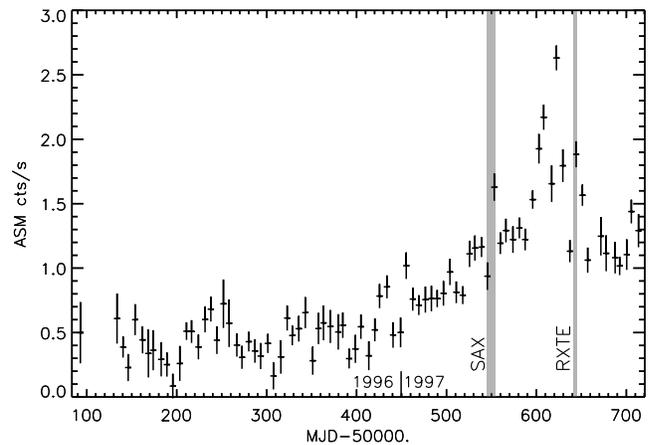,width=8.8truecm}}}
\caption[]{\label{ASM} RXTE ASM light curve of Mrk 501 during 1996 and 1997
with the epochs of pointed observations with SAX and RXTE  marked by the 
shaded bars.} 
\end{figure}

Markarian 501 is the second closest (z=0.034) BL Lacertae object. 
From its quiescent state radio to X-ray energy distribution
($\alpha_{\rm rx} \approx 0.62$, $f_\nu \propto \nu^{-\alpha}$) it    
can be described as intermediate between 
X-ray bright BL Lacs (XBLs) and radio bright objects (RBLs).
The object usually displays a relatively steep
X-ray spectrum (e.g. $\alpha_x=1.77$ observed by ROSAT in 1991)
indicating that the energy loss dominated high energy part of the
synchrotron spectrum is observed in X-rays 
(Lamer et al. \cite{Lamer96}).  X-ray spectra with  
$\alpha_x$ significantly less than unity, as required for the maximum
of the synchrotron power output to be at hard X-rays, have not been measured 
during earlier observations (1975-1990, Ciliegi et al. \cite{Ciliegi}).    
The 1 keV flux densities measured in these observations vary by a factor
of $\sim 3$.

\begin{table}[!h]
\caption[]{\label{obs_log} Observation log}
\begin{tabular}{lllrr}\hline
\noalign{\smallskip}
ObsID & start & end & \multicolumn{2}{c}{exposure [sec]} \\
      &\multicolumn{2}{c}{[MJD-50000.0]} & PCA & HEXTE$^a$\\    
\noalign{\smallskip}\hline
\noalign{\smallskip}
20421-01-01-01 & 640.9746  & 640.9916   & 1456  &  873 \\
20421-01-01-00 & 641.1764  & 641.1916   & 1296  & 1709 \\
20421-01-02-01 & 641.9740  & 641.9918   & 1520  &  640 \\
20421-01-02-00 & 642.1519  & 642.1919   & 2688  & 1771 \\
20421-01-03-01 & 642.9733  & 642.9918   & 1584  &  971 \\
20421-01-03-00 & 643.1520  & 643.1920   & 2880  & 1843 \\
20421-01-04-01 & 643.9526  & 643.9922   & 1616  & 1078 \\
20421-01-04-00 & 644.1526  & 644.1923   & 3152  &  996 \\
20421-01-05-01 & 644.9530  & 644.9922   & 1744  & 1129 \\
20421-01-05-00 & 645.1530  & 645.1922   & 3376  & 2135 \\
\noalign{\smallskip}\hline
\noalign{\smallskip}
\multicolumn{5}{l}{$^a$ added exposures of clusters A and B} 
\end{tabular}
\end{table}

\begin{table*}[htb]
\caption[]{\label{spec} Results of broken power law fits to the individual spectra}
\begin{tabular}{lllllllllll}\hline
\noalign{\smallskip}
MJD & duration & $\alpha_1$& $\pm (1\sigma)$ & $E_{\rm break}$& $\pm (1\sigma)$ & 
$\alpha_2$& $\pm (1\sigma)$ & 1 keV norm & $\pm (1\sigma)$  & $\chi^2$ (d.o.f)\\ 
& [days] & \multicolumn{2}{c}{($E<E_{\rm break}$)} &
\multicolumn{2}{c}{[keV]}  & \multicolumn{2}{c}{($E>E_{\rm break}$)} &
\multicolumn{2}{c}{${\rm [(cm^2\; s\; keV})^{-1}]$} & \\      
\noalign{\smallskip}\hline
\noalign{\smallskip}
50640.9831 &0.0170&0.756&0.015 &5.519&0.149 &1.054&0.008 &0.1702&0.0030&1.03 (224)\\ 
50641.1840 &0.0152&0.767&0.013 &5.984&0.186 &1.028&0.009 &0.1666&0.0030&1.04 (224)\\ 
50641.9829 &0.0178&0.664&0.016 &5.380&0.148 &0.962&0.007 &0.1400&0.0030&1.04 (224)\\ 
50642.1719 &0.0400&0.687&0.010 &5.856&0.143 &0.922&0.006 &0.1421&0.0020&0.95 (224) \\ 
50642.9826 &0.0185&0.653&0.014 &5.621&0.189 &0.889&0.008 &0.1306&0.0026&0.96 (224) \\ 
50643.1720 &0.0400&0.669&0.011 &5.604&0.152 &0.893&0.006 &0.1227&0.0020&1.11 (224) \\ 
50643.9724 &0.0396&0.674&0.015 &5.546&0.161 &0.944&0.007 &0.1381&0.0030&0.98 (224) \\ 
50644.1725 &0.0396&0.722&0.009 &5.889&0.131 &0.956&0.005 &0.1555&0.0018&0.98 (224) \\ 
50644.9726 &0.0393&0.746&0.010 &5.981&0.140 &1.029&0.007 &0.1803&0.0032&0.97 (224) \\ 
50645.1726 &0.0393&0.730&0.008 &5.724&0.090 &1.035&0.005 &0.1780&0.0020&1.02 (224) \\ 
\noalign{\smallskip}\hline
\noalign{\smallskip}
\multicolumn{2}{l}{integral spectrum}
                  &0.702&0.019 &5.576&0.208 &0.960&0.005 &0.1510&0.0040&0.98$^a$ (224)\\
\noalign{\smallskip}\hline
\noalign{\smallskip}
\multicolumn{11}{l}{$^a$ 1\% systematic error allowed}\\
\end{tabular}
\end{table*}

In 1995 Mrk 501 was detected as a source of
TeV gamma-rays by the {\em Whipple} team (Quinn et al. \cite{Quinn})
for the first time. Its flux level corresponded to  8\% of that of 
the Crab Nebula. The detection was confirmed by the HEGRA
collaboration (Bradbury et al. \cite{Brad}).
A recent compilation of TeV observations of Mrk 501 
(Protheroe et al. \cite{Protheroe}) shows 
the source at very high level with flares up to 10 Crab throughout
the entire observing season of 1997 well into the July observations
presented here. 

A similar brightening of Mrk 501 during 1997 has been observed 
in X-rays by the  RXTE All Sky Monitor (ASM), the peak 
brightness in June 1997 being $\sim$ 40 mCrab (Fig. \ref{ASM}).
The TeV and ASM (2-10 keV) light curves show similar flaring
activity during 1997, the correlated X-ray and VHE gamma-ray 
variability will be discussed in a forthcoming paper.

At the onset of this high activity in April 1997 observations with the
SAX satellite were carried out (Pian et al. \cite{Pian}).
They find Mrk 501 with an unusually hard X-ray spectrum,
extending up to photon energies of 100 keV with the
hardest spectrum observed on 16 April, when the source
was brightest.

\section{Observations and data analysis}
\label{data}

The bright X-ray state of Mrk 501 in June/July 1997 triggered 
intense multi-frequency observations from radio to VHE gam\-ma-rays.  
Here we report on RXTE observations between 11 July and 16 July 1997
resulting in 21300 seconds of good data in 10 pointings 
(Tab. \ref{obs_log}).

We have used {\tt ftools 4.0} for the reduction of the PCA and HEXTE
data. 
PCA good times have been selected from the standard 2 mode
data sets using the following
criteria: target elevation $> 10^\circ$, pointing offset $<0.01^\circ$,
and all 5 PCUs switched on. 
For the resulting intervals spectra and light curves were extracted using 
all Xenon layers. The corresponding PCA background spectra and light curves 
were derived with  {\tt pca\-backest 1.5}.

After applying the above elevation and offset criteria, HEX\-TE 
spectra and light curves have been binned and background subtracted 
using the off-source looking intervals.

Spectral fitting was performed with {\tt XSPEC 10.0} using the latest detector 
response matrix from 26 August 1997 for PCA  and the 
response files from
20 March 1997 for HEXTE. 
The PCA spectra in the PHA channel range 3-60 (2.5-27 keV)
and the spectra from both HEXTE  clusters in the 
channel range 15-100 (15-100 keV) were combined for each
of the 10 pointings and then 
fitted with single and broken power law spectra. The uncertainty
in the absolute flux calibration of the HEXTE instrument,
mainly due to uncorrected dead-time effects  
(see Rothschild et al. \cite{Roth}), was accounted for
by allowing for a scaling factor between the PCA and HEXTE spectra. 
Setting PCA to 1.0, the mean best fit value for the HEXTE 
normalization was 0.65.

The absorbing column density on the line of sight to Mrk 501
was fixed to $2.87 \cdot 10^{20} {\rm cm^{-2}}$ as derived from
the ROSAT spectrum (Lamer et al. \cite{Lamer96}). 
This value is larger than the galactic value as derived from
$21\;{\rm cm}$ H I measurements ($1.73 \cdot 10^{20} {\rm cm^{-2}}$,
Stark et al. \cite{Stark}). However, the effect of this difference 
on the spectral results in the RXTE energy range is negligible. 
Note that Pian et al. \cite{Pian} used the galactic H I column density
\nh=$1.73\cdot 10^{20} {\rm cm^{-2}}$   
for the spectral fits to the SAX data and that the uncertainty in the \nh 
value towards the source may have a significant effect on the spectral indices
measured with the LECS instrument at the softest X-ray energies.

\section{Spectra and variability}

Single power law models with  $N_H$ fixed to any reasonable value
do not yield acceptable fits to the data of any pointing, whereas  
broken power law models with break energies between 5.5 and 6.0 keV
give excellent fits to the individual spectra (see Tab. \ref{spec}).  
Below the break point the model spectra are exceptionally hard with
energy indices ranging from
0.65 to 0.77. In fact an X-ray synchrotron spectrum of this hardness 
has never been observed in any other XBL (Ciliegi et al. 
\cite{Ciliegi}, Lamer at al. \cite{Lamer96}). 
The spectrum measured on 16 April 1997 by  
SAX in about the same energy band (2.14-10 keV) is even slightly harder 
($\alpha=0.59$, Pian et al. \cite{Pian}). As the PCA has no 
sensitivity below 2 keV, we are not able to verify the  break
at $\sim 2\; {\rm keV}$ in the SAX spectrum.

\begin{figure}[htb]
\par{\centerline{\psfig{figure=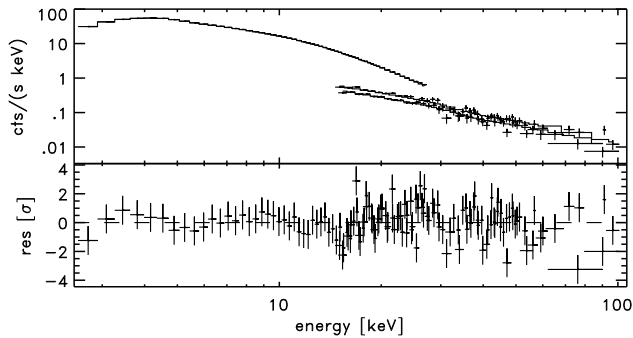,width=8.8truecm}}}
\caption[]{\label{allspec} Top panel: PCA and HEXTE countrate spectra of the
entire observation fitted with a broken power law model (see Tab. \ref{spec}
for the model parameters). Bottom panel: Fit residuals with 1\% systematic 
uncertainty allowed. } 
\end{figure}

Above the break point the PCA spectrum steepens by up to 0.3 in energy index.
In some of the spectra $\alpha_2$ exceeds unity, indicating that the maximum
of the synchrotron power output is reached in the hard X-ray range.
However, there is no  evidence for further steepening of the spectrum up to
100 keV.

During  the RXTE observations the PCA count-rate varied by $\sim 30\%$,  
but in contrast to the results from the April 1997 SAX observations  
an anti-correlation of flux and spectral
hardness was observed both below and above the spectral break. 
Such an anti-correlation
has never been reported in Mrk 501 and is quite unusual
for X-ray BL Lac objects altogether. Pian et al. found a spectral 
hardening by $\Delta \alpha = 0.32$ in the SAX MECS 2-10 keV band during an
increase of the flux in this energy band by a factor of 2.4 .

\begin{figure}
\par{\centerline{\psfig{figure=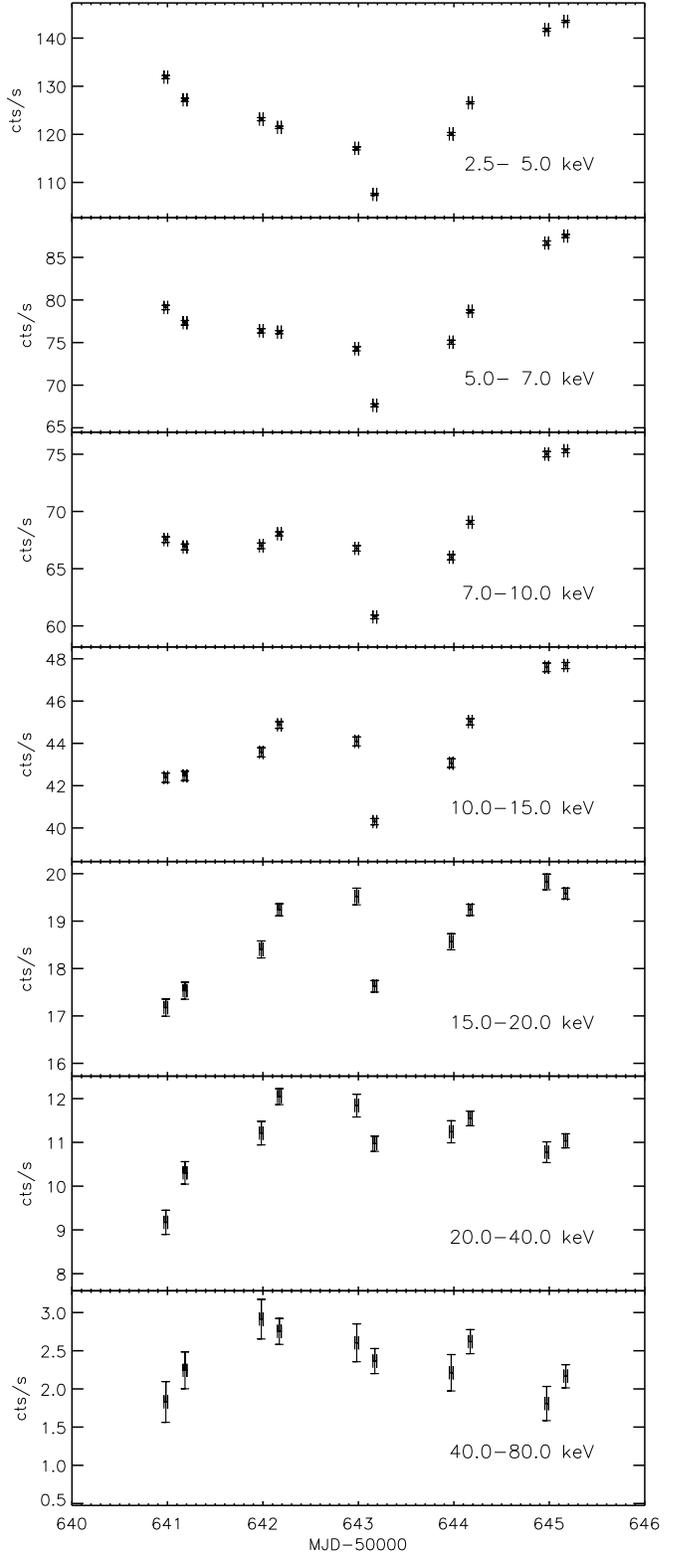,width=8.8truecm}}}
\caption[]{\label{pca_lc} Background subtracted 
PCA light curves in different energy bands.} 
\end{figure}

\begin{figure}
\par{\centerline{\psfig{figure=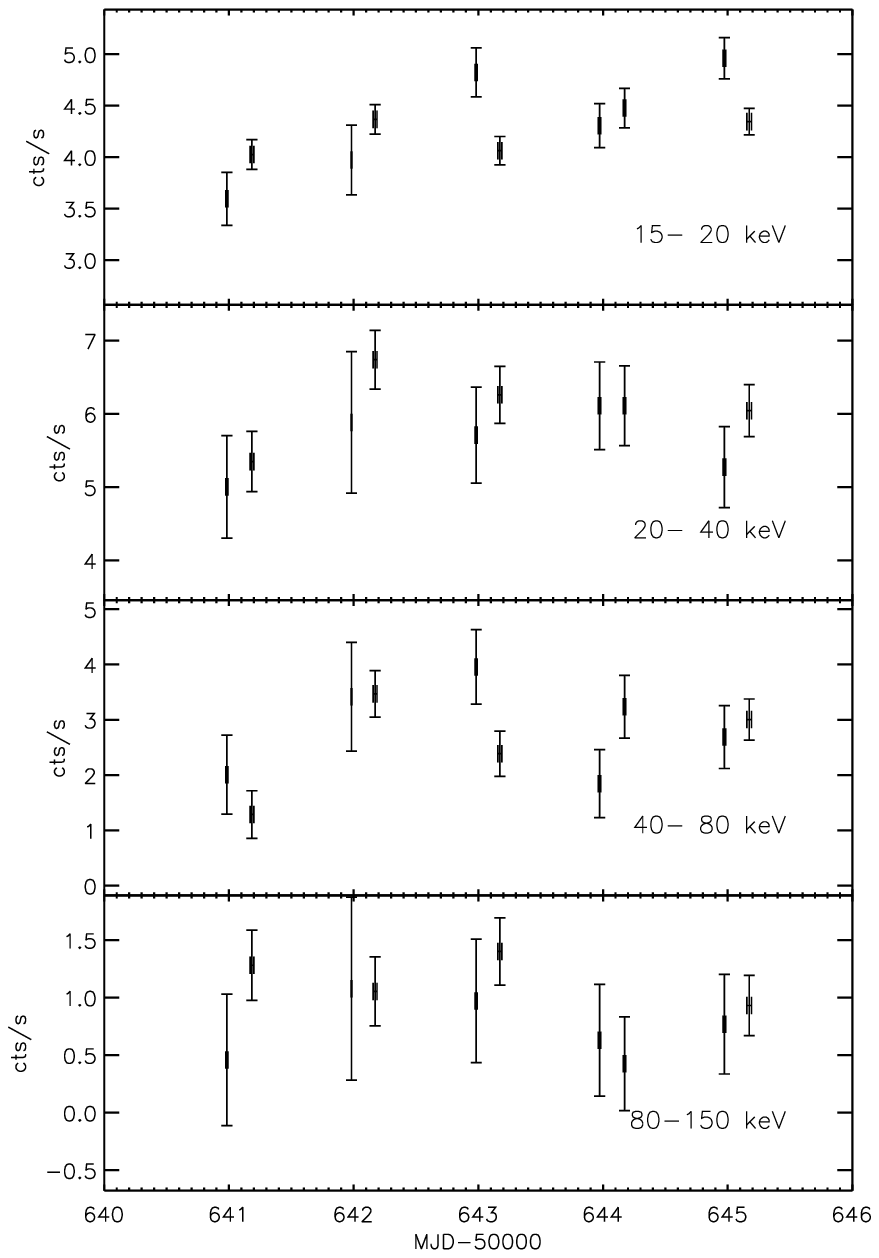,width=8.8truecm}}}
\caption[]{\label{hexte_lc} Background subtracted 
HEXTE light curves in different energy bands, the count-rates of
clusters A and B have been added. Note the good agreement of the PCA and HEXTE 
light curves in the 3 energy bands between 15 and 80 keV.} 
\end{figure}

The large degree of spectral variability and its anti-corre\-lat\-ion
with total flux implies great differences between the light curves
in the individual X-ray energy bands. 
We hence derived light curves for seven and four different channel 
ranges of the PCA and HEXTE detectors, respectively 
(Figs. \ref{pca_lc} and \ref{hexte_lc}).
The light curves in the softer bands show a decline by 20\% 
over 3 days with a well defined  minimum at MJD 50643.2 and
a subsequent increase by 30\% during the following 2 days.
On the other hand the hardest bands are dominated by a feature with a 
broad maximum around MJD 50642. 
A smooth transition with superpositions of both morphologies 
is observed in the intermediate energy bands.  
The independent variability of soft and hard X-rays requires the existence
of at least two emission components which is not evident from the 
X-ray spectrum alone.    

The sharpest feature in the 2-20 keV lightcurves  is a decrease
in flux by 10\% within 4.5 hours. In an EXOSAT ME lightcurve (0.7-8 keV,
Giommi et al. \cite{Giommi}) similar timescales and amplitudes have 
been found. In an investgation of the spectral variability of 6 BL Lac
objects Giommi et al. found tight correlations of 
flux and spectral hardness for any of the objects, including Mrk 501. 
This highlights the peculiarity of the spectral variability reported here.

\section{Conclusions}

Our RXTE observations from July 1997 show that the period of
increased X-ray brightness continued throughout 1997 and was
not limited to a short flare in April. The object clearly was
in a long-lasting high-state in 1997 rather than exhibiting
a short, spectacular X-ray flare during the epoch of the SAX
observations.
We confirm the extraordinary hard X-ray spectrum which extends 
up to 100 keV. The peak of the synchrotron emission is observed
in the hard X-rays, more than 2 orders of magnitude higher than
in earlier observations (e.g. by ROSAT in 1991, Lamer et al. \cite{Lamer96}).

During the RXTE observations in July the total flux and the
spectral hardness show an anti-correlation. 
This is a very unusual spectral behaviour for any BL Lac object, it has 
not been observed previously in Mrk 501 and is in marked
contrast to the fact that the SAX observations of Pian et al.
showed a flatter spectrum during the brighter stage (there
are only two different brightness and spectral levels in the SAX
data set rather than a well-established correlation).
The flux-spectral index relations should hence not be regarded
as universal and viable models should explain both kinds of
behaviour.

The broken power law (or gradual steepening of the spectrum) is
consistent with synchrotron cooling of a single component and
does not require the superposition of different particle
distributions. A homogeneous jet and a magnetic field of $0.4 {\rm G}$ 
are capable of producing the integrated spectral signature.
We have found rather dramatic spectral differences in the light 
curves which cannot be explained by a single population but
requires the contributions of at least two, if not more spectral 
components. This illustrates the great importance of well-sampled
monitoring with instruments covering a wide spectral range.

\begin{acknowledgements} 

We thank J. Swank for accepting the request for RXTE TOO
observations of Mrk 501, and the RXTE schedulers for the
flexible observation planning.
D. Gruber and W. Heindl are acknowledged for their assistance 
with the HEXTE data analysis. We thank R. Staubert for a careful reading
of the manuscript. 
RXTE ASM data are provided by the ASM/RXTE teams at MIT and
GSFC. PCA and HEXTE data were obtained through the HEASARC Online Service,
provided by NASA/GSFC.
This work was supported by DARA under grant 50 OR 96031,
and the DFG (SFB 328).

\end{acknowledgements}

\end{document}